\documentclass[conference]{IEEEtran}
\ifCLASSINFOpdf
\else
\fi
\usepackage{amsthm}
\usepackage{cite}
\usepackage{balance}
\usepackage{latexsym}
\usepackage{graphicx}
\usepackage{caption}
\usepackage{soul,color}
\usepackage{amsmath}
\usepackage{algorithm}
\usepackage{authblk}
\usepackage[noend]{algpseudocode}
\usepackage[utf8]{inputenc}
\usepackage[T1]{fontenc}
\theoremstyle{definition}

\theoremstyle{definition}

\begin{document}
%
\title{On the Foundation of NOMA and its Application to 5G Cellular Networks}


\author[1, 4]{\small Hikmet Sari}
\author[2]{\small Ali Maatouk}
\author[3]{\small Ersoy Caliskan}
\author[2]{\small Mohamad Assaad}
\author[3]{\small Mutlu Koca}
\author[1]{\small Guan Gui}

\affil[1]{ NUPT, 66 Xinmofan Road, Gulou District, Nanjing, 210003 China}
\affil[2]{ CentraleSup\'elec, Plateau de Moulon, 91192 Gif sur Yvette, France}
\affil[3]{ Bogazici University, Electrical \& Electronics Eng. Dept., Bebek 34342 Istanbul, Turkey}
\affil[4]{ Sequans Communications, 15 – 55 Boulevard Charles de Gaulle, 92700 Colombes, France}

\maketitle

\begin{abstract}
	Non-Orthogonal Multiple Access (NOMA) is recognized today as a most promising technology for future 5G cellular networks and a large number of papers have been published on the subject over the past few years. Interestingly, none of these authors seems to be aware that the foundation of NOMA actually dates back to the year 2000, when a series of papers introduced and investigated multiple access schemes using two sets of orthogonal signal waveforms and iterative interference cancellation at the receiver. The purpose of this paper is to shed light on that early literature and to describe a practical scheme based on that concept, which is particularly attractive for Machine-Type Communications (MTC) in future 5G cellular networks. Using this approach, NOMA appears as a convenient extension of orthogonal multiple access rather than a strictly competing technology, and most important of all, the power imbalance between the transmitted user signals that is required to make the receiver work in other NOMA schemes is not required here.
\end{abstract}

\IEEEpeerreviewmaketitle

\section{INTRODUCTION}
	Non-Orthogonal Multiple Access (NOMA) is currently a hot research topic for the physical layer of future 5G cellular networks, and more particularly for Machine-Type Communications (MTC) in that context. The interest in this multiple access technique originated from a well-established result in multi-user information theory, which says that orthogonal multiple access is not optimal in general and that superposition coding coupled with successive interference cancellation (SIC) provides an optimal solution for multiple access \cite{No1}, \cite{No2}.  

	Historically, time-division multiple access (TDMA) and frequency-division multiple access (FDMA) have been known and used in various forms for quite a long time. Focusing on digital cellular networks, the two major standards in second-generation (2G) cellular networks were the Global Standard for Mobile Communications (GSM) and IS-95. The first one of these was based on TDMA, and the second was based on code-division multiple access (CDMA) \cite{viterbi1995cdma}. For 3G networks, the winner was the CDMA technology, and the so-called Wideband CDMA (WCDMA) became the standard \cite{holma2007wcdma}. All of these networks were based on single-carrier transmission. Finally, 4G networks were based on the multicarrier transmission technology known as orthogonal frequency-division multiplexing (OFDM), previously used for terrestrial digital video broadcasting (DVB-T), WiFi, and WiMAX. In terms of multiple access, WiFi continued to use TDMA, but WiMAX used orthogonal frequency-division multiple access (OFDMA) which uses the frequency dimension of OFDM for resource allocation \cite{No3}. As for the 3GPP Long-Term Evolution (LTE) and LTE-Advanced standards \cite{No4} and \cite{holma2011lte}, they used OFDMA on the downlink and single-carrier FDMA (SC-FDMA) on the uplink in order to reduce the peak-to-average power ratio (PAPR) of the transmitted signal. 

	All of these multiple access techniques are orthogonal and ensure that no interference exists between users in ideal conditions. In TDMA only one user is active at a time, and in conventional FDMA only one user is active at a given frequency. In CDMA orthogonality is ensured by the properties of the Walsh-Hadamard (WH) sequences used for signal spreading. Finally, although individual user signals overlap in frequency in the case of OFDMA, orthogonality is achieved thanks to the carrier spacing of 1/T, where T is the symbol period. Of course, in all of these techniques, orthogonality on the uplink requires perfect synchronization between different user signals.

	Until the development of multi-user information theory, orthogonality of different user signals was always perceived as a most desirable property. But analysis of the channel capacity which indicated that orthogonal multiple access is not always optimal opened up new perspectives and research directions for future networks. Recently, a large number of papers have been published on non-orthogonal multiple access (NOMA), which is perceived as a most promising technology for 5G cellular networks (See, e.g., \cite{No5, No6, No7, No8}). The analysis in these papers promise substantial gains compared to conventional orthogonal multiple access. The purpose of the present paper is to put into context the current work on NOMA, quantify its potential, and point out previous work on the subject \cite{No9, No10, No11, No12}, which seems to be unnoticed by current researchers. Using that approach, NOMA appears as a convenient extension of orthogonal multiple access rather than a strictly competing technology.

	The paper is organized as follows: In the next section, we recall the basic principle of NOMA and we quantify the gain that this technique can achieve with respect to orthogonal multiple access in different scenarios. In Section III, we review the earlier work on NOMA that is based on the joint use of two orthogonal multiple access schemes. Using this concept, we describe in Section IV a practical NOMA scheme which elegantly combines OFDMA and Multi-Carrier CDMA (MC-CDMA) and turns out to be particularly suitable for accommodating two user groups with different profiles in terms of data rate requirements. Finally, we give simulation results in Section V and our conclusions in Section VI.

\section{PRINCIPLE OF NOMA}
	To describe the basic principle of NOMA, we will focus here on a two-user uplink channel in a cellular network. We will assume that user 1 has a strong signal power $P_1$ and user 2 has a weaker signal power $P_2$. In these conditions, the receiver can detect the user-1 signal in the presence of interference from the user-2 signal, and then it can subtract the detected user-1 signal from the received signal to detect the weaker user-2 signal without interference. Assuming that the channel is an additive white Gaussian noise (AWGN) channel of normalized bandwidth $W = 1$ Hz, the user 1 capacity in bits per Hertz is given by 
	\begin{equation}
		R_1=log_2\big(1+ \frac{P_1}{P_2 + N_0} \big)
	\end{equation}
	where $N_0$ is the noise spectral density (the noise power in the unit bandwidth $W = 1$ Hz). After detecting the signal transmitted by user 1, the receiver can subtract this from the received signal and detect the user 2 signal in the absence of interference. The user 2 capacity is given by
	\begin{equation}
		R_2=log_2\big(1+ \frac{P_2}{N_0} \big)
	\end{equation}
	and consequently, the total capacity for the two users is expressed as
	\begin{equation}
		R=R_1+R_2=log_2\big(1+ \frac{P_1}{P_2 + N_0} \big)+log_2\big(1+ \frac{P_2}{N_0} \big).
	\end{equation}
	A simple manipulation of this equation shows that
	\begin{equation}
	R=log_2\Big((1+ \frac{P_1}{P_2 + N_0} )(1+ \frac{P_2}{N_0} )\Big)=log_2\big(1+ \frac{P}{N_0} \big)
	\end{equation}
	where $P=P_1+P_2$ is the total power. This indicates that the capacity of the multi-user channel is identical to that of a single-user channel with the same total power. 
	
	The situation is actually not different for orthogonal waveform multiple access (OWMA). Without any loss of generality, consider an OFDMA scheme with 2 users. Again, $P_1$ will designate here the power of the user-1 signal, $P_2$ will designate the power of the user-2 signal, and $P=P_1+P_2$ is the total power. We write $P_1= \alpha P$ and $P_2=(1-\alpha)P$, with $0\leq \alpha \leq 1$. The signal power being uniformly distributed over the N carriers composing the OFDMA signal, the bandwidth allocation to the two users follows the same proportions as the signal power. In other words, we have $W_1=\alpha W$ and $W_2=(1-\alpha)W$, where bandwidth $W_1$ is allocated to user 1 and $W_2$ is allocated to user 2. The capacity equations for the two users are given by
	\begin{equation}
	R_1=\alpha log_2\big(1+ \frac{P_1}{W_1  N_0} \big)=\alpha log_2\big(1+ \frac{P}{N_0} \big)
	\end{equation}
	and
	\begin{equation}
	R_2=(1-\alpha)log_2\big(1+ \frac{P_2}{W_2  N_0} \big)=(1-\alpha) log_2\big(1+ \frac{P}{N_0} \big)
	\end{equation}
	respectively. The total capacity $R=R_1+R_2$ is therefore identical to the NOMA capacity given by equation (4). In summary, when the user signals do not have relative attenuations, both OWMA and NOMA achieve the single-user channel capacity, and the two multiple access technique do not have any difference in terms of capacity.
	
	The difference between the two multiple access techniques appears when one of the user signals is subject to a different attenuation than the other signal. Suppose that user 2 signal is attenuated by $6$ dB while the user 1 signal has no attenuation. In that case, the OFDMA capacity becomes 
	\begin{equation}
	R_{OFDMA}=\alpha log_2\big(1+ \frac{P}{N_0} \big)+(1-\alpha) log_2\big(1+ \frac{P/4}{N_0} \big)
	\end{equation}
	and the NOMA capacity will be
	\begin{equation}
	\begin{split}
	&R_{NOMA}=log_2\Big(1+ \frac{\alpha P}{(1-\alpha)P/4+N_0} \Big) \\
	&+log_2\Big(1+ \frac{(1-\alpha)P/4}{N_0} \Big) =log_2\Big(1+ \frac{(1+3 \alpha)P}{4 N_0}  \Big).	
	\end{split}
	\end{equation}
	To compare these two capacities, assume now that $\alpha=0.8$ and $ P  / N_0 = 15$ so that the single-user channel capacity is 4 bits per Hertz. In that case, (7) will read $R_{OFDMA}=3.65$ and (8) will read $R_{NOMA}=3.78$. Comparing these numbers, we can see that NOMA increases the two-user channel capacity by $3.5\%$ in this particular case. Pursuing the comparison further, it turns out that the advantage of NOMA increases when the parameter $\alpha$ is reduced and when the attenuation of the user-2 signal is further increased. But the capacity increase offered by NOMA does not come completely for free. Since user signals interfere with each other, iterative detection with interference cancellation is needed. When one signal is significantly weaker than the other, the strong signal can be detected with a small penalty and then subtracted from the received signal to detect the weak signal. Next, the weak signal is subtracted from the received signal to make more reliable second iteration decisions on the symbols of the strong signal, and so forth, until performance gets close to that of interference-free transmission. This process works fine when there is a strong imbalance between the two user signals, but it will have convergence problems when the two signals have similar powers.
\section{EARLY LITERATURE}
	The interest in NOMA today is closely related to the emergence of research projects on 5G cellular networks. The main research topics for the definition of the physical layer of 5G networks have been Massive MIMO, Waveform Design, and Millimeter-Wave Technologies, and multiple access is a key component of waveform design. Most papers on NOMA have been published over the past few years, and interestingly, the authors seem to be completely unaware of a series of papers which laid the foundation of NOMA over 15 years ago. This early literature did not use the word NOMA, but the principle of NOMA and all of the ingredients which characterize this technique were disclosed back in the year 2000. Below is a short summary of this work whose basic principle is to use two sets of orthogonal signal waveforms. We describe it here focusing on the combination of TDMA and OCDMA as in \cite{No9}, where the TDMA signal set is used in full and OCDMA signal set is used in part.
	
	Consider a simple TDMA system with $N$ users in which each user gets one data symbol per frame of $N$ symbols. The multiple access channel has a bandwidth of $NW$ Hz, where $W$ is the bandwidth which would be required to transmit the signal of the individual users if they transmitted alone. This scheme thus accommodates $N$ users without any interference. To accommodate additional users (say $M$ users, where $M < N$), a second signal set is used. The second signal set too is an orthogonal set, but the two sets are not mutually orthogonal. Specifically, the second set used here is the OCDMA signal set formed of length-$N$ Walsh-Hadamard (WH) sequences. In the resulting multiple access scheme with $N + M$ users, the first $N$ users do not interfere with each other, and the same applies to the second set of $M$ users, but each user from the first set interferes with every user from the second set, and vice versa. That is, we have here a NOMA scheme with interference between two groups of users, and iterative detection with serial interference cancellation is needed to detect the transmitted symbols.
	
\begin{figure}[h!]
	\centering
		\includegraphics[width=88mm]{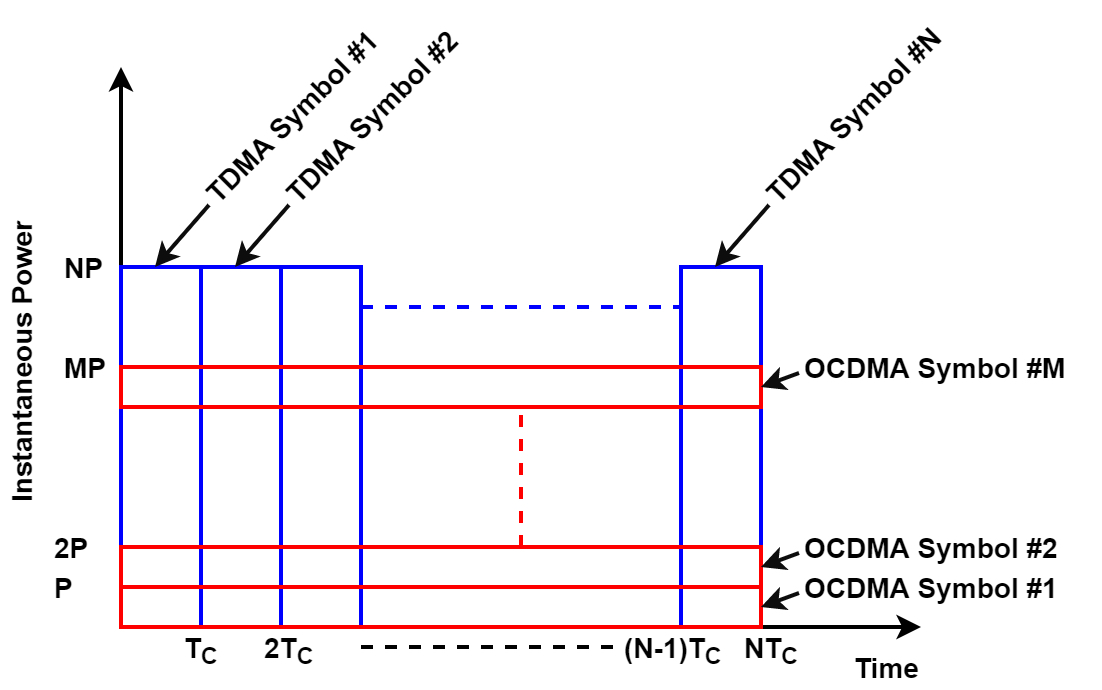}
		\caption{An illustration of the combined TDMA/OCDMA scheme, where the TDMA signal set is used in full and the OCDMA signal set is used partially.}
		\label{fig:demo}
	\end{figure}

	The basic principle is schetched in Fig.\ref{fig:demo}, which shows how $M$ OCDMA symbols are stacked to a block of $N$ TDMA symbols. On the abscissa of this figure, $T_C$ designates the OCDMA chip duration as well as the TDMA symbol duration, and $T=N \cdot T_C$ is the OCDMA symbol duration, which is also the duration of the TDMA symbol block. The transmitted instantaneous power is $P$ for each OCDMA symbol and $N \cdot P$ for each TDMA symbol so that both TDMA and OCDMA symbols have an energy of $E=N\cdot P\cdot T_C=P\cdot T$. The TDMA signal set here is used in full, while the OCDMA signal set is used only partially. This picture clearly shows that preliminary decisions can be made on the TDMA symbols as long as $M$ remains small compared to $N$.

	Let us write down the equations describing the transmitted signal. The time index describing the symbol position in a TDMA block is denoted $n$. Symbol $a_n$ with $1\leq n \leq N$ is assigned to TDMA user $ \# n$. The symbols assigned to the OCDMA users are denoted $b_m$ with $1\leq m\leq M$. We also write the WH sequences used for signal spreading as $W_m=(w_{m,1},w_{m,2},…..,w_{m,N} )$ for $m=1,2,….,M$. Using this notation, the transmitted signal can be written as
	\begin{equation}
	x_n=a_n + \frac{1}{\sqrt{N}}\sum_{m=1}^{M} w_{m,n}b_m
	\end{equation}
	for $ n=1,2,…,N $. The division by $ \sqrt{N} $ in (9) is to preserve the symbol energy during the symbol spreading process. The received signal can be written as $ r_n=x_n+u_n $ with $u_n$ being the additive noise for $ n=1,2,…,N $. Provided that the number of OCDMA users $ M $ is not too large, the interference term in (9) remains small compared to the TDMA symbol power, and the received signal sample $ r_n $ can be sent to a threshold detector to make a decision on the transmitted $ a_n $ symbol. Once these first-iteration decisions are made on all $ a_n $ symbols, the estimated symbol values are subtracted from the received signal samples according to $ y_n=r_n-\hat{a}_n $, where for each $ n $, $ \hat{a}_n $ stands for the decision made on symbol $a_n$. Referring back to (9), we can write $y_n$ as
	\begin{equation}
	y_n=a_n-\hat{a}_n + \frac{1}{\sqrt{N}} \sum_{m=1}^{M}w_{m,n}b_m + u_n.
	\end{equation}
	Assuming $\hat{a}_n= a_n$, (10) simplifies to
	\begin{equation}
	y_n=\frac{1}{\sqrt{N}} \sum_{m=1}^{M}w_{m,n}b_m + u_n.
	\end{equation}
	The next operation in the receiver is to perform signal despreading and make decisions on the OCDMA symbols. Signal despreading consists of
	\begin{equation}
	\begin{split}
	z_k&=\frac{1}{\sqrt{N}}\sum_{n=1}^{N}w_{k,n}y_n \\
	&= \frac{1}{\sqrt{N}}\sum_{n=1}^{N}w_{k,n} \bigg(\frac{1}{\sqrt{N}} \sum_{m=1}^{M}w_{m,n}b_m + u_n \bigg) \\
	&= b_k + \frac{1}{\sqrt{N}} \sum_{n=1}^{N}w_{k,n} u_n.
	\end{split}
	\end{equation}
	The second term is a noise term with identical variance to that of the original noise. The first iteration decisions on the OCDMA symbols are made by passing the $ z_k $’s to a threshold detector.
	
	Once the first-iteration decisions are also made for the $\{b_m,m=1,2,…,M\}$  symbols, their interference can be cancelled to make second-iteration decisions on the $\{a_n,n=1,2,…,N\}$ symbols. The process is as follows: For each n, compute $v_n=r_n-\frac{1}{\sqrt{N}}\sum_{m=1}^{M}w_{m,n}\hat{b}_m$, where for each $m$, $\hat{b}_m$ is the decision on $b_m$. Assuming that $\hat{b}_m= b_m$ for all $m$ and using (9), we get $v_n=a_n+u_n$. This signal is next sent to a threshold detector to make a decision on $a_n$ in the absence of interference. The second-iteration decisions are obviously more reliable than the first-iteration decisions, and the process continues as in the first iteration to make second-iteration decisions on the $\{b_m,m=1,2,…,M\}$ symbols. Additional iterations can further improve performance in some cases, but the results show that two iterations are sufficient when $M$ is small. 
	
	The concept described above is not just applicable to multiple access. It is equally applicable to single-user transmission, and therefore the terminology of “channel overloading” was used in \cite{No11} to describe it. The basic idea is that once the channel is fully loaded using an orthogonal signaling scheme (orthogonal transmission for a single-user channel or orthogonal multiple access for a multiuser channel), it is overloaded through the superposition of a second signal to the first one. Optimum joint detection being too complex to implement, the receiver in practice takes the form of an iterative receiver with interference cancellation. For multiple access, the recent NOMA literature focuses on the superposition of two user signals, but the concept reviewed in this section goes actually further and superposes the signals of two user groups.
\section{NOMA FOR 5G}
\begin{figure}[h!]
	\centering
		\includegraphics[width=88mm]{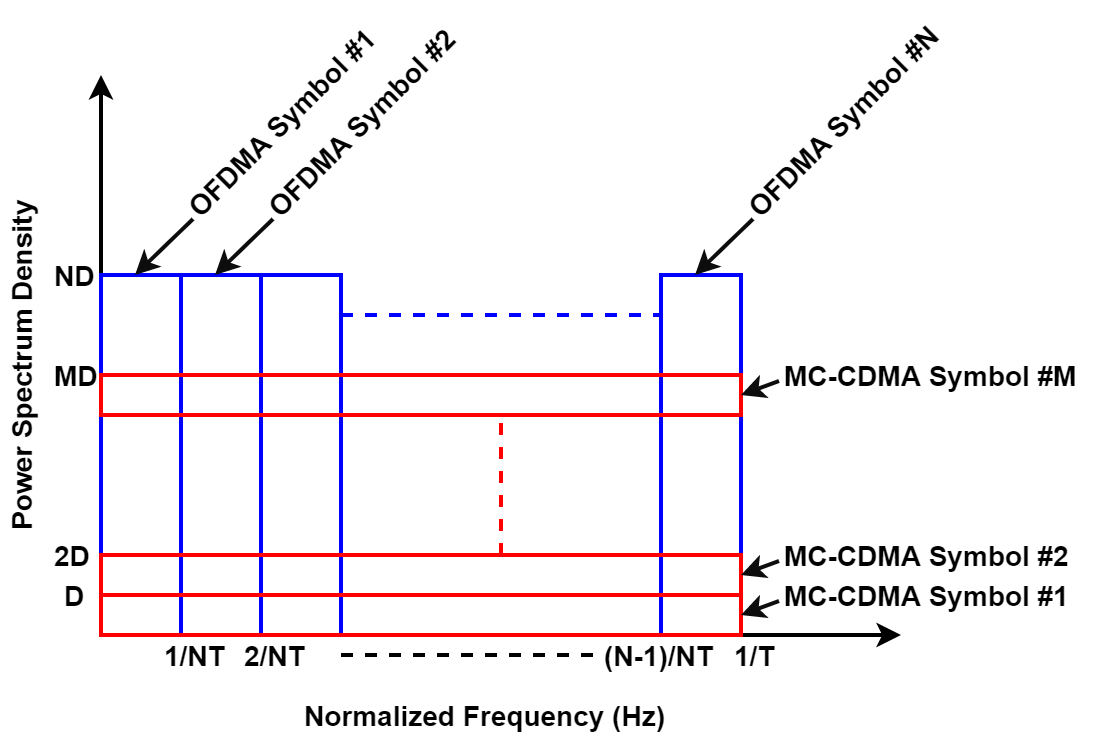}
		\caption{An illustration of the combined OFDMA/MC-CDMA scheme, where the OFDMA signal set is used in full and the MC-CDMA signal set is used partially.}
		\label{fig:demo1}
	\end{figure}

	Since OFDMA has been the basic multiple access scheme used in 4G cellular systems and it has also been adopted by the 3GPP for mobile broadband (eMBB) services in 5G \cite{techreport2}, we will now describe a frequency-domain NOMA scheme, which consists of using OFDMA for the first group of users and Multi-Carrier CDMA (MC-CDMA) for the second group. The principle is exactly the same as the one described in the previous section, with the frequency dimension substituted for the time dimension as shown in Fig. \ref{fig:demo1}. In this figure, $1/NT$ is the carrier spacing, the OFDMA symbols have a power spectrum density (PSD) of $N \cdot D$ Watt/Hz, and the MC-CDMA symbols that are superposed to them have a PSD of $D$ Watt/Hz.
	
\begin{figure}[h!]
		\includegraphics[width=90mm, height=21mm]{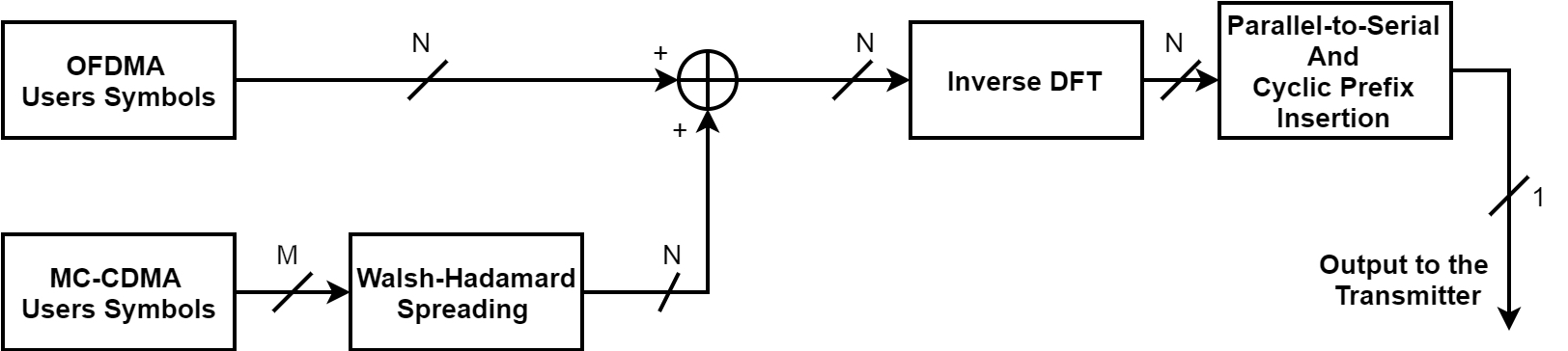}
		\caption{Transmitter block diagram of a NOMA scheme using OFDMA for the first set of users and MC-CDMA for the second set of users.}
		\label{fig:transmitter}
	\end{figure}

	To describe this NOMA technique further, consider an OFDMA system with $N$ carriers and without any loss of generality assume that each carrier is assigned to a separate user. Such a system accommodates $N$ users providing one QAM symbol to each of them during every OFDM symbol. Applying the concept described in Section III, we superpose to this OFDMA signal a set of MC-CDMA signals carrying information for a second group of users. The mathematical equations of Section III remain the same except that here, $n$ with $1\leq n \leq N$ designates the carrier index, and $x_n$ given by (9) designates the signal transmitted on the $n^{th}$ carrier. A simple block diagram of the transmitter is shown in Fig. \ref{fig:transmitter}.
	\\
	The output of the OFDMA Users block is an N-dimensional QAM symbol vector $\{a_n,n=1,2,…,N\}$ and the MC-CDMA Users block is an M-dimensional symbol vector $\{b_m,m=1,2,…,M\}$. The Walsh-Hadamard Spreading box spreads the MC-CDMA symbols over the $N$ carriers and outputs an N-dimensional vector that is summed with the OFDMA symbols vector. The resulting signal block is passed to an N-point inverse DFT operator followed by the insertion of cyclic prefix (CP) between consecutive inverse DFT blocks. 
	
\begin{figure}[h!]
		\includegraphics[width=88mm]{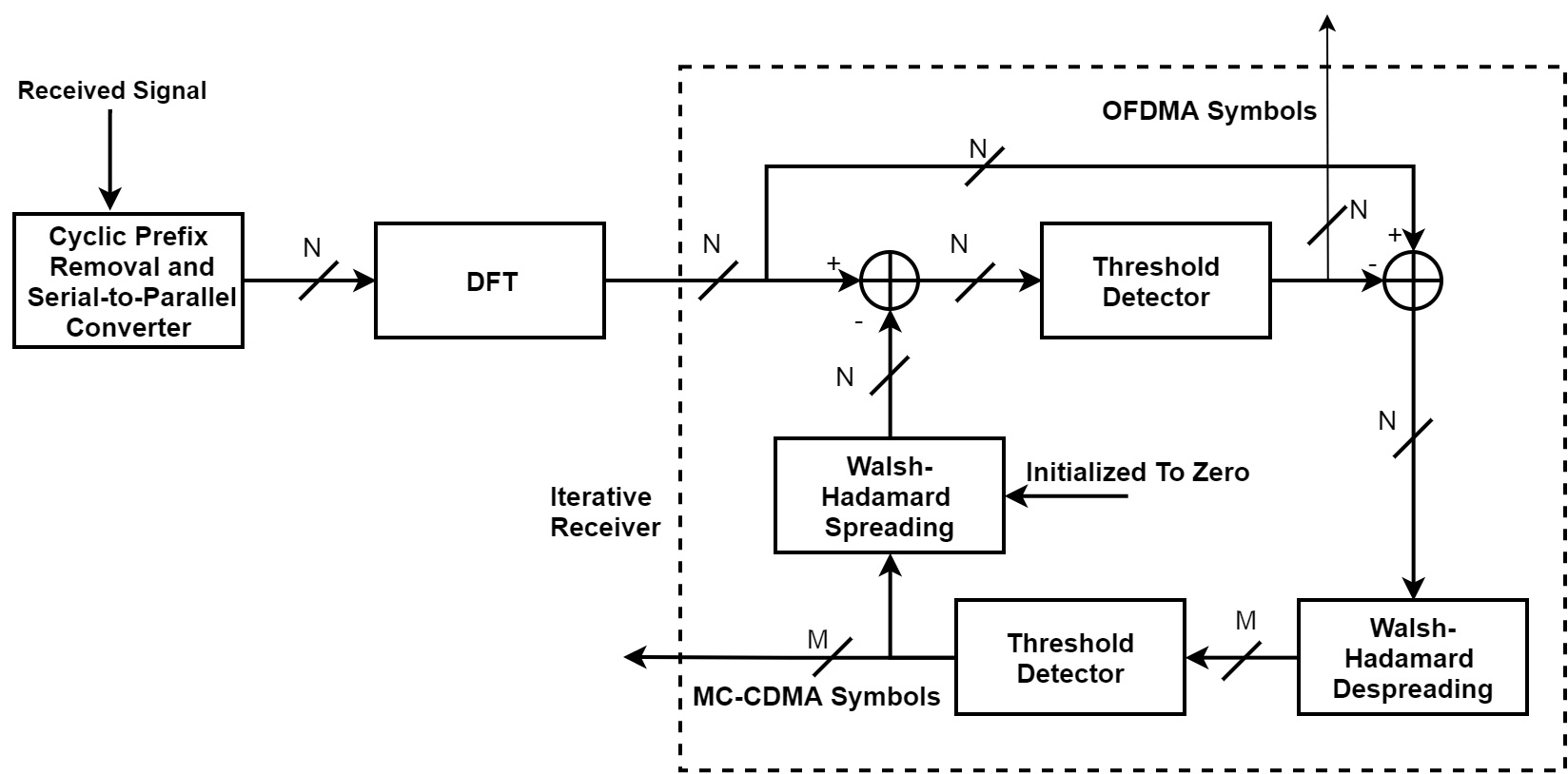}
		\caption{Receiver block diagram for the NOMA scheme of Fig. \ref{fig:transmitter}.}
		\label{fig:reciever}
	\vspace{-10pt}
	\end{figure}

The corresponding receiver is sketched in Fig. \ref{fig:reciever}. 
\\
After CP removal in the time domain, the signal is converted to the frequency domain by means of an N-point DFT. The output of this operator $\{r_n,n=1,2,…,N\}$ is directly passed to a threshold detector to make first-iteration decisions on the OFDMA symbols. These are denoted $\{\hat{a}_n,n=1,2,…,N\}$. These decisions are subtracted from the DFT operator output to provide $\{y_n=r_n-\hat{a}_n,n=1,2,...,N\}$ and this signal is passed to a Walsh-Hadamard despreader. The despreader output is next sent to a threshold detector to make first-iteration decisions on the MC-CDMA symbols $\{b_m,m=1,2,…,M\}$. These decisions, which are denoted $\{\hat{b}_m,m=1,2,…,M\}$, are Walsh-Hadamard spread, the spreader output block is subtracted from $\{r_n,n=1,2,…,N\}$, and the resulting signal is passed to a threshold detector to make second-iteration decisions on $\{a_n,n=1,2,…,N\}$. Finally, these decisions are subtracted from the threshold detector inputs, the resulting signal is Walsh-Hadamard despread and passed to a threshold detector to make second-iteration decisions on $\{b_m,m=1,2,…,M\}$. The process can continue to make further iterations as required, but two iterations are sufficient in practice for small values of $M$.

	At this point, it is important to discuss the number of MC-CDMA user signals which can be superposed to the OFDMA user signals without a significant performance degradation. The WH sequences used for signal spreading are binary sequences with components $\pm 1$. Due to the multiplicative term $1/\sqrt{N}$ used in signal spreading, the interference from each MC-CDMA user on OFDMA users is of the form $\pm 1/\sqrt{N}$. When the number of MC-CDMA users reaches $\sqrt{N}$, the peak interference amplitude reaches $1$ and the eye diagram of the OFDMA signal becomes closed. In this case, errors occur in the first-iteration decisions of OFDMA symbols even in the absence of noise, which means that the corresponding bit error rate (BER) curve features an error floor. Correspondingly, we limit for the moment the number of MC-CDMA users to $\sqrt{N}$, although this does not  represent a strict bound. Indeed, an iterative receiver employing soft decisions instead of hard decisions as described in \cite{No12} will help accommodating a higher number of MC-CDMA users.
\section{SIMULATION RESULTS}
	Performance of the NOMA scheme presented in the previous section was evaluated using computer simulations. The simulations were carried out over an additive white Gaussian noise (AWGN) channel using an OFDMA/MC-CDMA scheme with $N = 64$ carriers, 16-QAM modulation for the OFDMA users, and QPSK for the MC-CDMA users. In a first set of simulations, the number of MC-CDMA users was $M = 4$ and in a second set $M$ was increased to $8$ such that the superposition of the MC-CDMA signals to the OFDMA signal leads to a closed eye diagram.
	
\begin{figure}[h!]
	\centering
		\includegraphics[width=88mm]{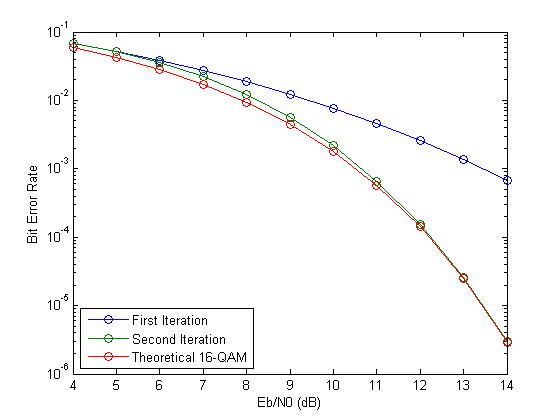}
		\caption{Bit error rate for OFDMA users for $N = 64$ and $M = 4$.}
		\label{fig:simulation1}
		\vspace{-20pt}
	\end{figure}

\begin{figure}[h!]
		\includegraphics[width=88mm]{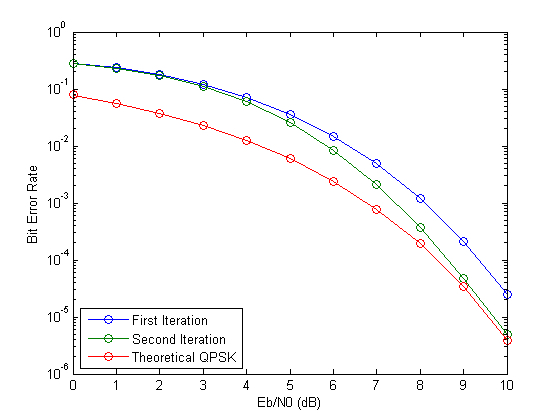}
		\caption{Bit error rate for MC-CDMA users for $N = 64$ and $M = 4$.}
		\label{fig:simulation2}
	\end{figure}

\begin{figure}[h!]
		\includegraphics[width=88mm]{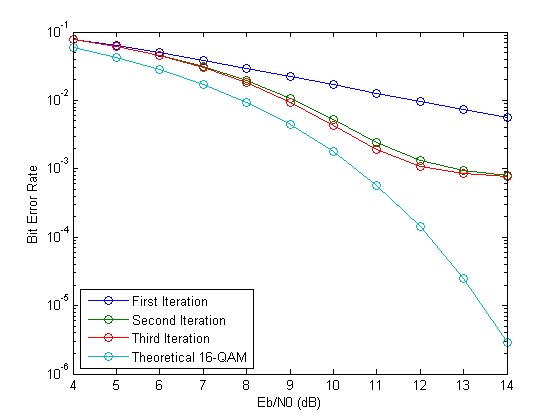}
		\caption{Bit error rate for OFDMA users for $N = 64$ and $M = 8$.}
		\label{fig:simulation3}
		\vspace{-25pt}
	\end{figure}

\begin{figure}[h!]
		\includegraphics[width=88mm]{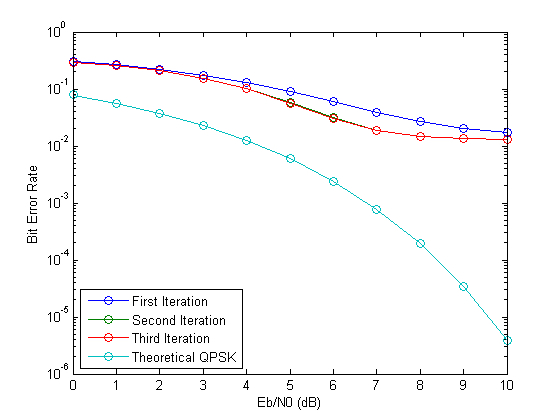}
		\caption{Bit error rate for MC-CDMA users for $N = 64$ and $M = 8$.}
		\label{fig:simulation4}
	\end{figure}

	With $M = 4$, the performance results are given in Fig. \ref{fig:simulation1} for OFDMA users and in Fig. \ref{fig:simulation2} for MC-CDMA users. As can be seen in Fig. \ref{fig:simulation1}, the BER curve of OFDMA users at the first iteration has a large gap from the theoretical BER curve of 16-QAM, but the second iteration gives a remarkable result and leads essentially to the same performance as the theoretical curve for BER values below $10^{-3}$. The gap at the first iteration can be explained by the level of interference from MC-CDMA users. Next, examining Fig. \ref{fig:simulation2}, we can see that performance of the MC-CDMA users has somewhat a different behavior. The performance curve at the first iteration has essentially the same shape as the theoretical BER curve of QPSK with a gap that is close to $1$ dB at BER values lower than $10^{-4}$. At the second iteration, performance improves sharply and the BER curve virtually coincides with the theoretical BER curve of QPSK at BER values below $10^{-5}$. These results indicate that in the case at hand no more than $2$ iterations are needed in the iterative receiver to cancel the interference between OFDMA users and MC-CDMA users. 
	\\
	The results corresponding to $M = 8$ are given in Fig. \ref{fig:simulation3} for OFDMA users and in Fig. \ref{fig:simulation4} for OCDMA users. Three iterations were made in this case. Fig. \ref{fig:simulation3} shows that at the first iteration, the BER curve decays only very slowly and reaches $5.10^{-3}$ at $E_b/N_0$ of $14$ dB. The second iteration reduces the bit error rate to some extent, but a BER floor slightly below $10^{-3}$ is clearly visible on that curve. A third iteration gave only a marginal improvement and did not change the value of the BER floor. Next, Fig. \ref{fig:simulation4} shows that the BER curve is very flat at the first iteration and that the second iteration provides a very small improvement. The third iteration provided no improvement at all and the BER floor is slightly higher than $10^{-2}$ on this figure. This result confirms that when hard decisions are made in the iterative receiver, the number of MC-CDMA users must be kept below $\sqrt{N}$ in order to avoid the appearance of a BER floor.
\section{CONCLUSIONS}
In this paper, we have addressed NOMA, which is a strong candidate today for machine-type communications in future 5G cellular systems. After describing the basic principle of this technique, we pointed out that its foundation actually goes back to the year 2000, a fact that seems to be unnoticed by authors of recent papers. The concept appeared in a series of papers published in that period describing multiple access using two orthogonal signal sets and iterative detection with serial interference cancellation. We first gave a comprehensive review of this technique using TDMA for the first set of users and OCDMA for the second set. Next, focusing on the context of 5G cellular systems, we described a practical NOMA scheme employing a combination of OFDMA and MC-CDMA, which can form an attractive solution for machine-type communications in 5G. In that approach, NOMA can be viewed as an extension of OFDMA to perform channel overloading and accommodate a higher number of users when all resources of OFDMA are used. Alternatively, OFDMA and MC-CDMA can be used to accommodate two user sets with different profiles and data rate requirements. The power imbalance which is required to make reliable detection in NOMA appears very naturally in this approach, because for identical symbol energy, the power spectrum density of MC-CDMA symbols is only $(1/N)^{th}$ of the OFDMA spectrum density.




\bibliographystyle{IEEEtran}
\bibliography{privacy}

\end{document}